# A New Combined European Permanent Network Station Coordinates Solution

**N. Panafidina, Z. Malkin**
Institute of Applied Astronomy of RAS, Kutuzova emb. 10, St. Petersburg, 191187, Russia

**R. Weber**
Institute of Geodesy and Geophysics, Vienna, Austria

**Abstract.** The EUREF (International Association of Geodesy (IAG) Reference Frame Sub-Commission for Europe) network of continuously operating GPS stations (EPN) was primarily established for reference frame maintenance, and also plays an important role for geodynamical research in Europe. The main goal of this paper is to obtain an independent homogeneous time-series of the EPN station coordinates, which is also available in SINEX format. A new combined solution of the EPN station coordinates was computed. The combination was performed independently for every week, in three steps: 1. the stated constraints on the coordinates were removed from the individual solutions of the Analysis Centers; 2. the deconstrained solutions were aligned to ITRF2000; 3. the resulting solutions were combined using the Helmert blocking technique. All the data from GPS week 900 to week 1302 (April 1997 – December 2004) were used. We investigated in detail the behavior of the transformation parameters aligning the new combined solution to ITRF2000. In general, the time-series of the transformation parameters show a good stability in time although small systematic effects can be seen, most likely caused by station instabilities. A comparison of the new combined solution to the official EUREF weekly combined solution is also presented.

## 1. Introduction

The EUREF permanent GPS network (EPN, http://www.epncb.oma.be/) is a network of continuously operating GPS stations, primarily established for the maintenance of the European Terrestrial Reference System ETRS89 and its successive realizations (Ádam et al. 2002). It is coordinated by the International Association of Geodesy (IAG) Regional Reference Frames Sub-commission for Europe (EUREF). Also, the EPN is a European densification of the global tracking network operated by the International Global Navigation Satellite Systems Service (IGS) (Beutler et al. 1999; Ferland et al. 2002).

The stations of the EPN are divided into a set of sub-networks so that each station is present in at least three sub-networks. Observations from each sub-network are processed by one Analysis Center (AC). Currently, 16 ACs are operating:

ASI - Telespazio S.p.A., Space Geodesy Center, Italy,
BEK - International Commission for Global Geodesy of the Bavarian Academy of Sciences, Germany,
BKG - Bundesamt für Kartografie und Geodäsie, Germany,
CODE - Center for Orbit Determination in Europe, Astronomical Institute of the University of Berne, Switzerland,

DEO - Delft Institute for Earth-Orientated Space Research, Delft University of Technology, the Netherlands,
GOP - Geodetic Observatory Pecny, Czech Republic,
IGE - Instituto Geografico Nacional, Spain,
IGN - Institut Geographique National, France,
LPT - Bundesamt fuer Landestopographie, Switzerland,
NKG - Nordic Geodetic Commission,
OLG - Observatory Lustbuehel Graz, Austria,
ROB - Royal Observatory of Belgium,
SGO - FOMI Satellite Geodetic Observatory, Hungary,
SUT - Slovak University of Technology Bratislava, Slovakia,
UPA - Universita di Padova, Italy,
WUT - Warsaw University of Technology, Poland.

All ACs except DEO and ASI use the Bernese software (Hugentobler et al. 2005) based on least-squares adjustment; DEO uses GIPSY software, and ASI uses MicroCosm software. Detailed information concerning data processing at the ACs is available on the EPN Central Bureau (CB) web-site (http://www.epncb.oma.be/_dataproducts/analysiscentres/index.html). Although EPN ACs are, in principal, free to choose an appropriate processing strategy, the EPN CB provides guidelines for some processing options that are recommended or mandatory for all the EPN ACs (http://www.epncb.oma.be/_organisation/guidelines/guidelines_analysis_centres.html#processingoptions). In particular, mandatory options include antenna phase center correction, observations' cut off angle, elevation-dependent weighting of observations, using IGS final orbits and Earth rotation parameters (ERPs), etc. As to the latter, it should be mentioned that before GPS week 1130, ACs were free to use either IGS or CODE orbits, which may cause some inconsistency between individual solutions.

The official EUREF weekly combined solution for station coordinates is one of the products of the EPN (the so-called ITRS series). It is currently computed at the BKG, Germany (before GPS week 1020, it was computed at CODE, Switzerland) and provides weekly coordinate solutions for about 160 stations in SINEX format (Habrich 2002). A regional network (such as EPN) solution can be expressed in the International Terrestrial Reference Frame (ITRF) using two main approaches: constraining coordinates of a set of selected fiducial stations to the values given in the ITRF (hereafter referred to as fiducial approach), and aligning a regional solution to the ITRF using Helmert transformation (Altamimi 2002).

Until GPS week 1302, the official EUREF solution was computed using the fiducial approach, where coordinates of some selected stations (8-14 for different periods) were tightly constrained (practically fixed) to the ITRF values (changing with time from ITRF94 to ITRF2000). Unfortunately, this solution is not suitable for high-precision geodetic applications because the approach used for the data processing, coupled with successive changes of the reference frames (ITRF versions), can cause jumps in coordinates and distortions of the network. In addition, for the constrained (fiducial) stations, the EUREF solution only gives linear movements defined by the ITRF coordinates and velocities. It should be mentioned that since GPS week 1303, the EPN Combination Center uses the minimal constraints approach for the weekly combination, which overcomes the deficiencies mentioned above.

There is another solution available at the EPN CB, intended to be used for geokinematics and based on the minimal constraints approach, the so-called "CLEANED Time Series" (Kenyeres et al. 2002; Kenyeres and Bruyninx 2004). This coordinate time-series seems to be much more homogeneous, but unfortunately it is not available in SINEX format (http://tau.fesg.tu-muenchen.de/~iers/web/sinex/format.php). Therefore, the main goal of this study was to obtain homogeneous coordinate time-series for EPN stations available in SINEX format for further geokinematic investigations.

At the first stage, all available weekly individual solutions for GPS weeks 900-1302 (April 1997 – December 2004) were used for the combination, except for the solutions from the DEO and ASI ACs because they do not provide statistical information (such as variance factors, number of unknowns and number of observations) needed for removing constraints from the solutions. The variance factor is also needed at the combination stage to compute the inverse normal system matrix.

The combination was done in three steps: first, each individual solution was de-constrained using a priori information provided in SINEX files. Then, the de-constrained solutions were transformed to ITRF2000 (Altamimi et al. 2002; Boucher et al. 2004) using a seven-parameter Helmert transformation. Finally, all individual solutions were combined using the Helmert blocking technique.

## 2. Combination strategy

### 2.1. Removing constraints

Most ACs provide constrained coordinate solutions, where coordinates of one or several stations are tightly constrained to the values given in the current ITRF. Before combining the individual solutions, these stated constraints have to be removed.

The constrained solution can be presented as a combination of two solutions: the a priori solution and the free solution (Brockmann 1996). The combination of two solutions can be written as:

$$\hat{\boldsymbol{\beta}}_c = \hat{\boldsymbol{\beta}}_1 + \boldsymbol{\Sigma}_1(\boldsymbol{\Sigma}_1 + \boldsymbol{\Sigma}_2)^{-1}(\hat{\boldsymbol{\beta}}_2 - \hat{\boldsymbol{\beta}}_1) \quad (1)$$

$$\boldsymbol{\Sigma}_c = \boldsymbol{\Sigma}_1 - \boldsymbol{\Sigma}_1(\boldsymbol{\Sigma}_1 + \boldsymbol{\Sigma}_2)^{-1}\boldsymbol{\Sigma}_1 \quad (2)$$

where $\hat{\boldsymbol{\beta}}_c$ and $\boldsymbol{\Sigma}_c$ are the combined solution and its covariance matrix, $\hat{\boldsymbol{\beta}}_1$, $\hat{\boldsymbol{\beta}}_2$, $\boldsymbol{\Sigma}_1$, $\boldsymbol{\Sigma}_2$ are the input solutions and the respective covariance matrices.

ACs provide the coordinate solutions in SINEX format with the a priori and statistical information. For the constraints removal, we consider $\hat{\boldsymbol{\beta}}_c$ and $\boldsymbol{\Sigma}_c$ as the constrained solution and its covariance matrix given in the SINEX-blocks SOLUTION/ESTIMATE and SOLUTION/MATRIX_ESTIMATE. For the first input solution and its covariance matrix, we take the a priori solution ($\hat{\boldsymbol{\beta}}_1 = \hat{\boldsymbol{\beta}}_{apr}$) and the a priori matrix ($\boldsymbol{\Sigma}_1 = \boldsymbol{\Sigma}_{apr}$) from the SINEX-blocks SOLUTION/APRIORI and SOLUTION/MATRIX_APRIORI. The second input solution is the free solution, which we want to obtain ($\hat{\boldsymbol{\beta}}_2 = \hat{\boldsymbol{\beta}}_{free}$, $\boldsymbol{\Sigma}_2 = \boldsymbol{\Sigma}_{free}$). Then, one can derive $\hat{\boldsymbol{\beta}}_{free}$ from Eqs. (1) and (2) as a combination of $\hat{\boldsymbol{\beta}}_c$ and $\hat{\boldsymbol{\beta}}_{apr}$. The result for the free solution and the covariance matrix is given by (Brockmann 1996):

$$\hat{\boldsymbol{\beta}}_{free} = \hat{\boldsymbol{\beta}}_{apr} + \boldsymbol{\Sigma}_{apr}(\boldsymbol{\Sigma}_{apr} - \boldsymbol{\Sigma}_c)^{-1}(\hat{\boldsymbol{\beta}}_c - \hat{\boldsymbol{\beta}}_{apr}) \quad (3)$$

$$\boldsymbol{\Sigma}_{free} = \left(\boldsymbol{\Sigma}_c^{-1} - \boldsymbol{\Sigma}_{apr}^{-1}\right)^{-1} \quad (4)$$

For each GPS week, all individual solutions were de-constrained using the formulas given above. Since usually only a few stations are constrained in a solution, the a priori covariance matrix was expanded to the dimension of the estimated covariance matrix using some large value (of the order of $10^4$) for the variances of the unconstrained stations, which corresponds to introducing a very loose constraint (about 100 m) for these stations.

The EUREF ACs use IGS Final orbits and ERPs in processing. The geocentre and scale are based on the satellite dynamics, and the rotations are determined by the used ERPs. Station coordinates taken directly from the de-constrained solutions may vary at a level of several decimeters. For this reason, we aligned all the de-constrained ACs solutions to the ITRF2000, as follows.

## 2.2. Aligning to ITRF2000

For the combination, all individual solutions have to be aligned to some common reference frame to avoid systematic differences between the solutions. This can be done in one step together with the estimation of the combined solution, or each individual solution can be aligned to some common reference frame before the combination. We used the second approach. After removing constraints, the obtained "free" individual solutions were aligned to ITFR2000 (ftp://lareg.ensg.ign.fr/pub/itrf/itrf2000/ITRF2000_GPS.SNX.gz) using a seven-parameter Helmert transformation.

Some stations with unreliable behaviour (Bruyninx et al. 2004; Stangl and Bruyninx 2005), and stations with poorly determined ITRF2000 coordinates and velocities (in total 18) were not used for the calculation of the transformation parameters. All remaining stations present both in ITRF2000 and in the individual solution were used with weights dependent on the position errors in both the AC solution and ITRF2000. For the latter, coordinates from the ITRF2000 catalog and their errors were propagated to the epoch of the individual solution using ITRF2000 positions and velocities and their errors. Stations for which the coordinate difference between the transformed individual solution and the propagated ITRF2000 position exceeded 3 cm were excluded and the transformation parameters were recomputed.

## 2.3. Combination

A combined solution was computed using the Helmert blocking method, which leads to the equation (Brockmann 1996):

$$\left(\sum_{i=1}^{m}\left(\Sigma_i^{-1}\right)\right)\hat{\boldsymbol{\beta}}_c = \sum_{i=1}^{m}\left(\Sigma_i^{-1}\hat{\boldsymbol{\beta}}_i\right) \qquad (5)$$

where $\boldsymbol{\beta}_c$ denotes the combined solution, $\Sigma_i$ is the inverse matrix of the normal equation system of an individual solution, $\boldsymbol{\beta}_i$ is an individual solution and $m$ is the number of individual solutions.

This simple superposition of normal equations is always possible if the individual observation series are independent. Unfortunately, this assumption is not exactly true for the individual solutions of various ACs because all of them use the same observations. Nevertheless, in this work they are considered as independent since correlations between individual ACs' solutions are not known and cannot be used.

For the variance of the unit weight of the combined solution, we get:

$$\hat{\sigma}_c^2 = \left(\sum_{i=1}^{m}\hat{\sigma}_i \cdot f_i + \sum_{i=1}^{m}(\hat{\boldsymbol{\beta}}_c - \hat{\boldsymbol{\beta}}_i)'\Sigma_i^{-1}(\hat{\boldsymbol{\beta}}_c - \hat{\boldsymbol{\beta}}_i)\right)/f_c \qquad (6)$$

where $f_i = n_i - u_i$ redundancy for the individual solution i

$n_i$           number of observations for the individual solution i

$u_i$           number of unknowns for the individual solution i

$f_c = n_c - u_c$ redundancy for the combined solution

$n_c = \sum_{i=1}^{m} n_i$    total number of observations

$u_c$           total number of unknowns

In order to make individual solutions within one week more homogeneous, each deconstrained solution was scaled before the combination so that the mean formal coordinate error for each individual solution became equal to the mean formal coordinate error for all individual solutions of a given week. No additional weighting was applied. We investigated other potential methods of weighting based on the analysis of the formal errors, and on systematic and random differences between individual and combined solutions. Since none of them led to noticeable improvements in the resulting coordinate time-series and their respective errors, we kept equal weighting of the AC solutions within the combination process. A better weighting scheme using variance component estimation is planned in the future.

For outlier detection, we first calculated the coordinate differences between the combined solution and each individual solution, and then the RMS value corresponding to the series of coordinate differences of all ACs was computed. A station was rejected from an individual solution if the coordinate difference at least for one component exceeded three-sigma. In general, up to six stations were excluded from one individual solution.

## 3. Results of the computation

For each GPS-week from 900 to 1302, the new combined solution was obtained using the procedure described above. In order to compare the new combined solution with the official EUREF weekly combined solution, we computed transformation parameters for both solutions to ITRF2000. Since the new combined solution was calculated in ITRF2000, the transformation parameters should be stable, have no jumps and be close to zero.

Figures 1 to 7 show the time-series of transformation parameters for the new combined solution and the official EUREF combined solution w.r.t. ITRF2000. The time-series of all transformation parameters between the official EUREF combined solution and ITRF2000 show noticeable seasonal variations and jumps in week 1143 in the time-series of translations and scale. The seasonal variations might be caused by the fiducial approach used for the combination. The jump in the time-series of translations and scale corresponds to the change of ITRF97 to ITRF2000. The time-series of the scale parameter shows a jump of about ~2.5 ppb in week 1143 and noticeable seasonal variations (see Fig. 1b). The expected influence of the reference frame change from ITRF97 to ITRF2000 is about 1.6 ppb according to the official transformation at the GPS week 1143 (ftp://lareg.ensg.ign.fr/pub/itrf/ITRF.TP).

The time-series of transformation parameters between the new combined solution and ITRF2000 show a better stability compared to the official EUREF combined solution, although time-series of most of the parameters show slight systematic biases and seasonal variations. This should be mostly caused by the usage of some unstable stations for aligning individual AC's solutions to ITRF2000. The number of stations common in EPN and ITRF2000 is nearly constant (see Fig. 8), but some worsening for the later GPS weeks can be expected because of the deterioration of ITRF2000 coordinates with time.

Figure 9 demonstrates correlations between the transformation parameters for the new combined solution. The obvious strong correlation between $T_y$ and $R_x$, $T_z$ and $R_y$, and anti-correlation between $T_x$ and $R_y$, $T_y$ and $R_z$ can be explained by the relatively small area covered by the EUREF network. Strong correlations between the transformation parameters make any physical interpretation difficult, and the parameters are only valid strictly within the coverage area. These correlations are also visible in the plots of time-series of the transformation parameters (Figs. 1 to 7).

To illustrate the difference between the EUREF combined solution and the new one, we show as an example coordinate time-series for stations SVTL (Russia) (see Fig. 10) and GOPE (Czech Republic) (see Fig. 11).

## 4. Conclusion

A new time-series of weekly combined solutions for the coordinates of EPN stations for GPS weeks 900-1302 has been computed. In this solution, tight constraints on the fiducial stations were removed, and the solution was aligned to the ITRF2000 using Helmert blocking. The new solution shows a better stability with respect to ITRF2000 than the EUREF solution, the only one available from the EPN CB in SINEX format.

Our solution can be easily recomputed using ITRF2004, when available. This should substantially improve its quality since up to now only about half of the EPN stations are present in the ITRF2000, some of them with poorly determined velocity. Our weekly solutions in SINEX format are publicly available on the GPS AC at the Institute of Applied Astronomy web-site (http://www.ipa.nw.ru/PAGE/DEPFUND/GEO/ac_gps/). The new solution is also submitted to the EUREF Time Series Monitoring Special Project (Kenyeres et al. 2002).

In summary, the method used for relating the combined solution to ITRF2000 can provide reasonable results only in case that the stations used for aligning the individual solutions to ITRF2000 are well distributed over the network, have well-determined ITRF coordinates and velocities, and are stable enough. Especially the selection of stable stations needs special consideration. Also, the procedure for removing constraints described above may be not sufficient if the constraints were too tight (Mareyen and Becker 2000). Nevertheless, as additional experimental computations have shown, this simplified procedure can provide very similar results to a minimally constrained solution in sense of random noise in station coordinate time-series.

## Acknowledgements

The authors are grateful to Jim Ray and three anonymous reviewers for their valuable comments and suggestions, which allowed us to improve the manuscript.

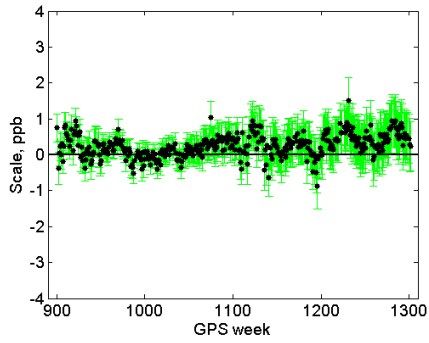
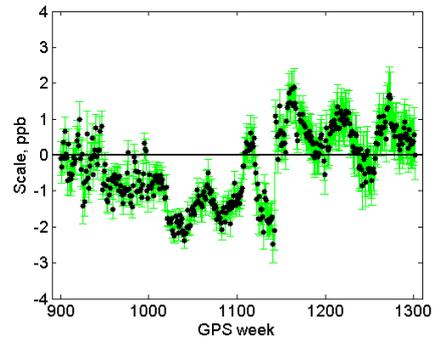

Figure 1a. Scale (ppb): new solution w.r.t. ITRF2000.

Fig 1b. Scale (ppb): official EUREF solution w.r.t. ITRF2000.

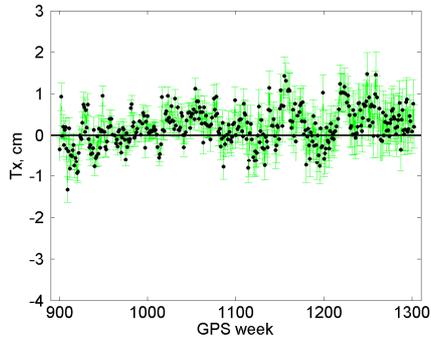
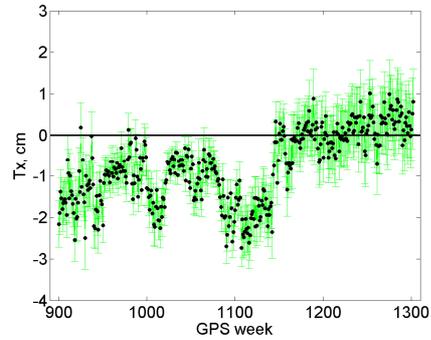

Figure 2a. X-translation (cm): new solution w.r.t. ITRF2000.

Figure 2b. X-translation (cm): official EUREF solution w.r.t. ITRF2000.

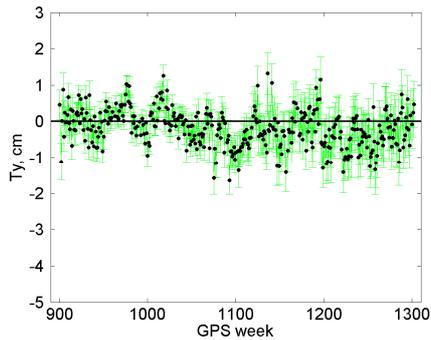
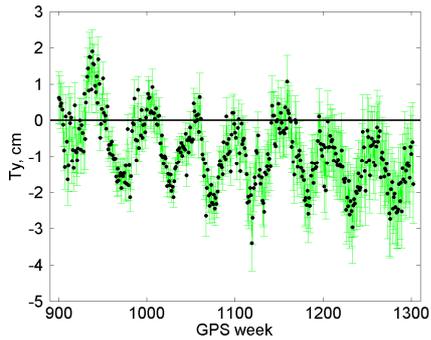

Figure 3a. Y-translation (cm): new solution w.r.t. ITRF2000.

Figure 3b. Y-translation (cm): official EUREF solution w.r.t. ITRF2000.

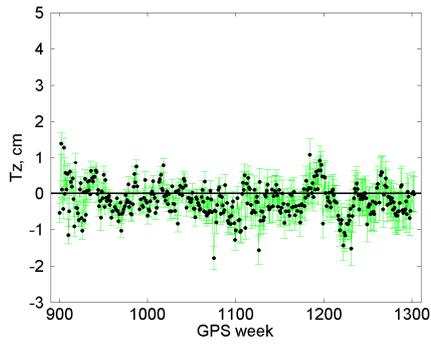
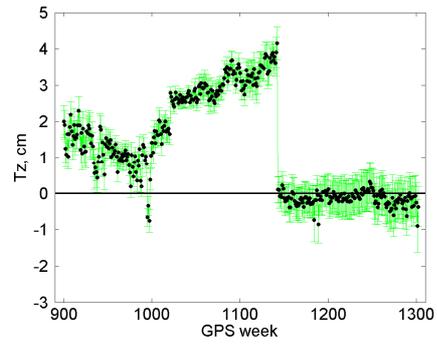

Figure 4a. Z-translation (cm): new solution w.r.t. ITRF2000.

Figure 4b. Z-translation (cm): official EUREF solution w.r.t. ITRF2000. The jump in week 1143 is caused by the change of the terrestrial reference frame.

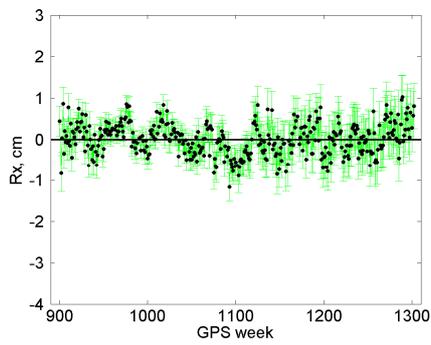
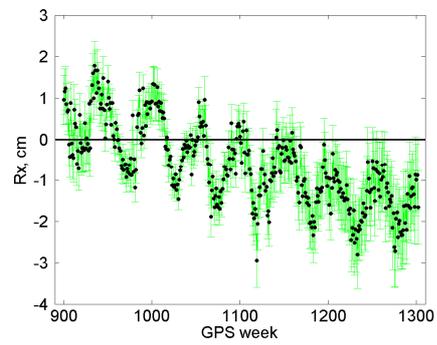

Figure 5a. X-rotation (cm): new solution w.r.t. ITRF2000.

Figure 5b. X-rotation (cm): official EUREF solution w.r.t. ITRF2000.

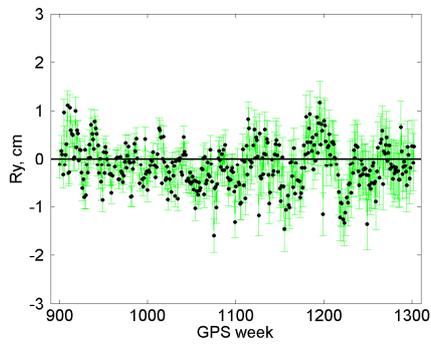
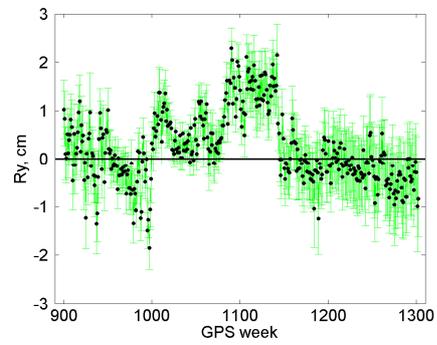

Figure 6a. Y-rotation (cm): new solution w.r.t. ITRF2000.

Figure 6b. Y-rotation (cm): official EUREF solution w.r.t. ITRF2000.

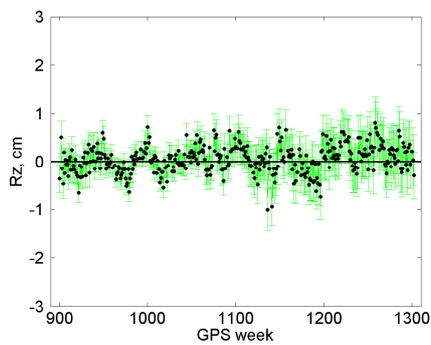
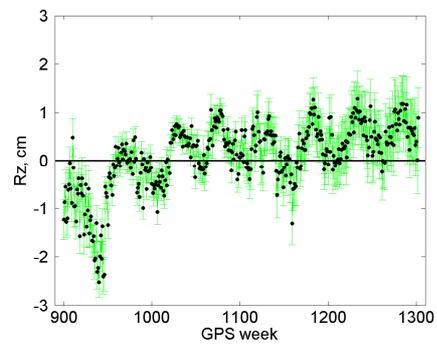

Figure 7a. Z-rotation (cm): new solution w.r.t. ITRF2000.

Figure 7b. Z-rotation (cm): official EUREF solution w.r.t. ITRF2000.

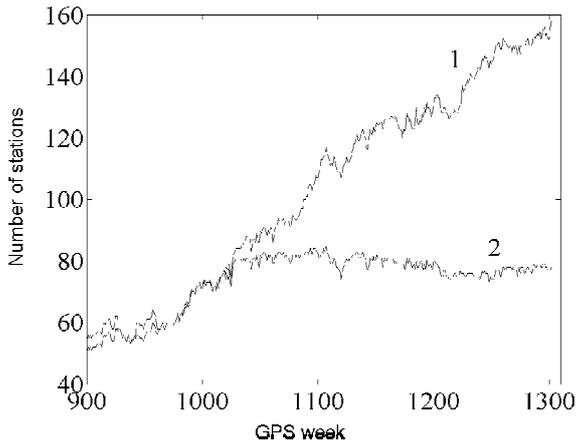

Figure 8. Number of EUREF stations (1), and number of stations common in the EUREF network and the ITRF2000 (2).

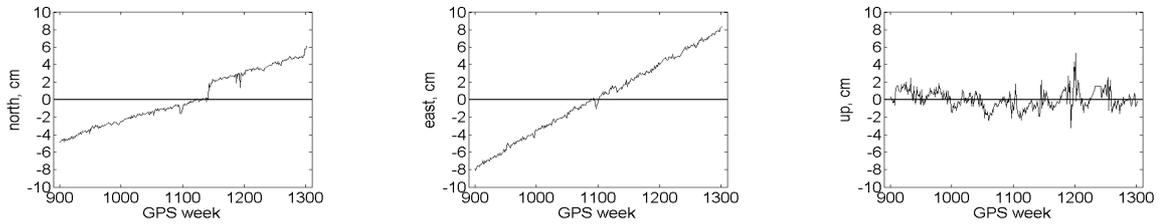

Figure 9. Correlation matrix: 1227 w.r.t. ITRF2000, new combined solution of week translations Tx, Ty, Tz, scale D, rotations Rx, Ry, Rz.

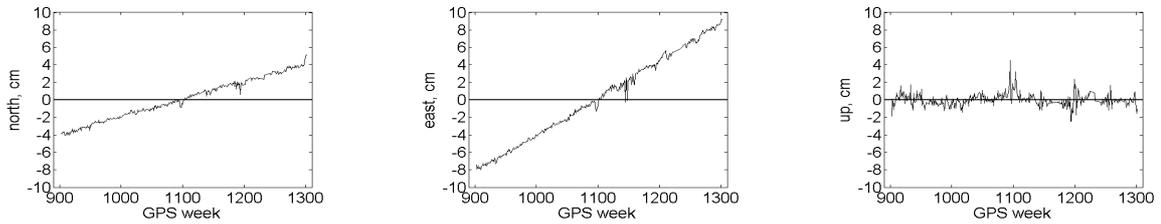

Fig 10a. SVTL, official EUREF combined solution, north, east, up components (cm)

Figure 10b. SVTL, new combined solution, north, east, up components (cm)

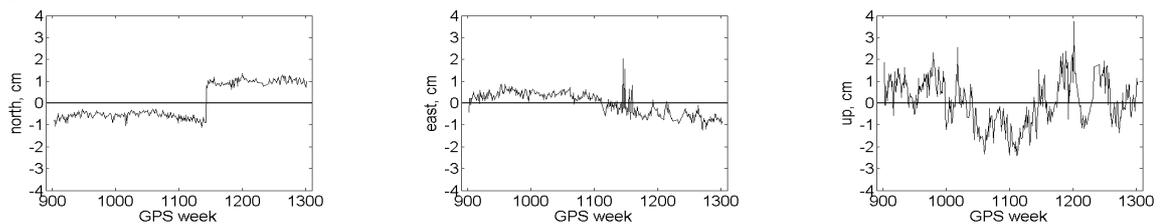

Fig 10c. SVTL, coordinate differences between the new combined solution and the official EUREF combined solution, north, east, up components (cm).

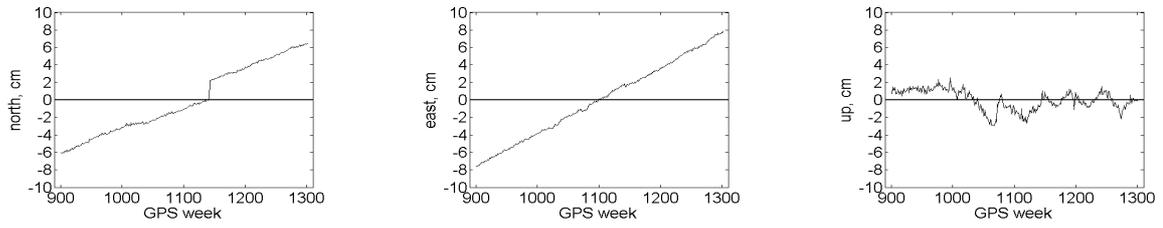

Fig 11a. GOPE, official EUREF combined solution, north, east, up components (cm)

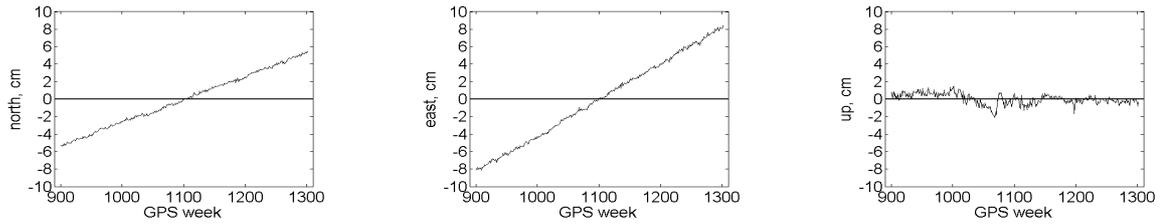

Figure 11b. GOPE, new combined solution, north, east, up components (cm)

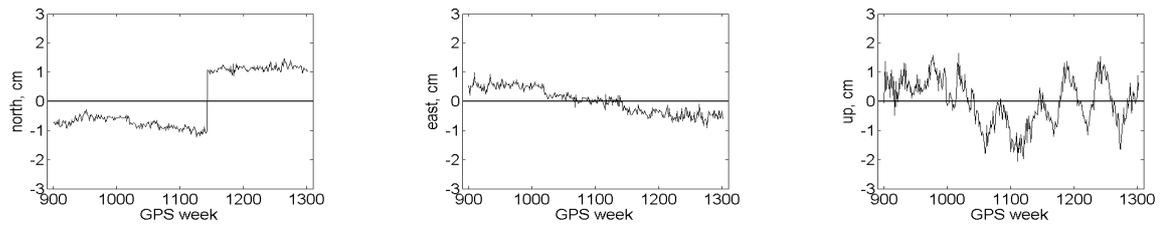

Fig 11c. GOPE, coordinate differences between the new combined solution and the official EUREF combined solution, north, east, up components (cm).